\documentstyle[epsfig,preprint,aps]{revtex}
\def\skip1{\vskip \baselineskip}
\def\skip2{\vskip 2\baselineskip}

\begin{document}

%Page0 (Title Page)

\newcommand{\bm}[1]{\mbox{\boldmath$#1$}}

\thispagestyle{empty}

\title{Zero-temperature TAP equations for the Ghatak-Sherrington model}

\author{F.A. da Costa $^{1,2}$ \ and J.M de Ara\'ujo $^{1,3}$ }

\address{
$^1$ Departamento de F\a'{\i}sica
Te\a'{o}rica e Experimental \\
Universidade Federal do Rio Grande do Norte \\
Campus Universit\a'{a}rio -- \ C.P. 1641 \\
59072-970 \hspace{8mm} Natal - RN, \hspace{8mm} Brazil \\
$^2$ Present address: Instituto de F\a'{\i}sica \\
Universidade Federal Fluminense \\
Av. Litor\^anea s/n -- Boa Viagem \\
24210-340 \hspace{8mm} Niter\'oi - RJ, \hspace{8mm} Brazil \\
$^3$ Departamento de Ci\^{e}ncias Naturais \\
Universidade Estadual do Rio Grande do Norte \\
59610-210 \hspace{8mm} Mossor\'o - RN, \hspace{8mm} Brazil\\ }

\maketitle

\newpage

%\tightenlines

\setcounter{page}{2}

\begin{center}{\large\bf Abstract}\end{center}
\vskip \baselineskip

The zero-temperature TAP equations for the spin-1 Ghatak-Sherrington
model are investigated. The spin-glass energy density (ground state)
is determined as a function of the anisotropy crystal field $D$ for
a large number of spins. This allows us to locate a first-order transition
between the spin-glass and paramagnetic phases within a good accuracy. The
total number of solutions is also determined as a function of $D$.

\vspace{12cm}

\begin{tabbing}
\=xxxxxxxxxxxxxxxxxx\= \kill
\>{\bf PACS Numbers:} \> 05.70.-a, 64.60.-i, 75.10.-b \\
\end{tabbing}

\newpage

Recently different authors have investigated some infinite-range
spin-glass models which display both continuous and first-order
transitions lines\cite{Rosenow96,Arenzon96,Sellitto97,Nogueira98,Rehker99,%
Schreiber99,Oppermann99,FeldmannOppermann99}.
In most cases, the first-order transition line starts at a tricritical point
and extends down to $T=0$. Whithin the replica approach, one should
in principle locate this line according to the program proposed by Parisi
\cite{ParisiA}. However, it is easy to see that this is a very hard
task, and has not been achieved so far. The difficulty has its origin in the
fact that at a first-order transition at least two phases with different
symmetries coexist with the same free energy but distinct order
parameters. In most spin-glass models at least one of these order
parameters must be obtained within an accurate numerical study of the
full Parisi treatment.

Another method to study disordered systems was introduced in order to
avoid replicas and is widely known as TAP approach \cite{TAP}. It is also
well known that this method presents another kind of numerical
difficulty, since the TAP equations have an exponentially large number of
solutions, most of them related to metastable states
\cite{DeDominicis80,BrayMoore80}. Nevertheless, it is possible to solve the
TAP equations for the Sherrigton-Kirkpatrick (SK) model \cite{SK} and get
some useful information on the nature of the spin-glass phase
\cite{NeTaka85,Nemoto87}.

Following SK, Ghatak and Sherrington \cite{GS} introduced a generalized
model to the case of integer spin $S_i=0,\pm1,\ldots,\pm S$ and
including a crystal-field term. This model is an example of the systems
mentioned in the first paragraph above. For $S=1$ it has a first-order
transition line that for some time was the object of some controversies
\cite{Lage82,Mottishaw85,daCosta94}. Although its location is presently
known within the replica-symmetric solution, it was not yet determined
from a full Parisi solution. Recently,  Feldmann and Oppermann \cite{Feldmann99}
showed that a fermionic Ising spin glass is equivalent
to the spin-1 Ghatak-Sherrington model.
These authors also considered a one-step replica-symmetry
breaking (1RSB) in order to locate the first-order transition line. For
$T=0$ they concluded that the transition is located at $D\approx0.881J$,
where $D$ represents the crystal field and $J^2/N$ the variance of the
random couplings, a result already known to one of us \cite{YokoiU}.
This should be compared to the replica-symmetric result which gives
$D\approx0.899J$ \cite{daCosta94}. In fact it results in a tedious algebra
but one can show that a second step in the replica-symmetry breaking (2RSB)
procedure gives $D\approx0.880J$ \cite{YokoiU}. Thus, one may suspect that a
full Parisi treatment would give a value for $D/J$ very close to $0.88$. It
is thus natural to look for an alternative treatment in order to check these
results.

The TAP equations for the Ghatak-Sherrington model were already obtained
a few years ago by Yokota\cite{Yokota92}. This author did an extensive
numerical study of the transition at a particular temperature, namely, $T=0.2J$.
He showed that the first-order transition should be located at
$D/J=0.85\pm0.05$. However, he did not search for the zero-temperature transition
which turns out to be simpler to analyse. On the other hand, the TAP equations
for the analogous fermionic Ising spin glass were recently obtained by Rehker
and Oppermann \cite{Rehker99}. Following \cite{Feldmann99} it is easy to see
that there is an exact mapping between the corresponding equations for both
models. Rehker and Oppermann \cite{Rehker99} have numerically studied
their equations for $T=0$ and found that the transition line is located
at $D \approx 0.8J$, without any mention to error bars. They also
obtained equations which determine the number of solutions for the
corresponding TAP equations at any temperature, but were unable to
solve them numerically even for $T=0$.

The purpose of the present note is twofold. Firstly, we show that carrying
out a numerical study with some reasonable numbers of spins, the TAP equations
for the Ghatak-Sherrington model can be used to determine the transition at
$T=0$ to a good accuracy. Secondly, we show that total number of such
solutions can be determined as a function of $D$ (or, equivalently, $\mu$ in
the fermionic-glass case \cite{Rehker99}).

We consider the Hamiltonian

\begin{equation}
H = -\sum_{(ij)}J_{ij}S_iS_j + D\sum_i S_i^2 ~,
\end{equation}

\noindent
where each spin $S_{i}$ \ ($i=1,2, \cdots , N$) can take the values
$-1,0$ and $1$ and the summations are over all distinct pairs $(i,j)$.
The random exchange couplings $J_{ij}$ have zero mean and variance
$J^2/N$. According to Eqs. (3--4) from Yokota \cite{Yokota92}, the TAP
equations for this system can be written as

\begin{eqnarray}
m_i&=& {\displaystyle\frac{2\sinh(\beta h_i)}{\exp(\beta \Delta_i)+
2\cosh(\beta h_i)}}  \nonumber  \\
p_i&=& {\displaystyle\frac{2\cosh(\beta h_i)}{\exp(\beta \Delta_i)+
2\cosh(\beta h_i)}}   \label{rsy}
\end{eqnarray}

\noindent
where

\begin{equation}
h_i = \sum_{j}J_{ij}m_j - \beta m_i\sum_j J_{ij}^2(p_j-m_j^2) ~,
\end{equation}

\noindent
and

\begin{equation}
\Delta_i = D-\frac{1}{2}\beta~\sum_j J_{ij}^2(p_j-m_j^2) ~,
\end{equation}
where $m_i$ and $p_i$ are thermal averages fo $S_i$ and $S_i^2$,
respectively.

At $T=0$ the above equation simplify to

\begin{eqnarray}
m_i &=& {\rm sgn}(h_i)\Theta(|h_i|-D) ~, \label{tap0} \\
h_i &=& \sum_{j}J_{ij}m_{j} , \nonumber
\end{eqnarray}
where $\Theta$ is the Heaviside step function. The energy density (in
units of $J$) is given by

\begin{equation}
f = -\frac{1}{N J}\sum_{(i,j)}J_{ij}m_im_j + \frac{D}{NJ}\sum_i p_i ~,
\end{equation}
where $p_i=m_i^2$ at $T=0$.

In the paramagnetic phase all $m_i$ are equal zero and so is its
energy density, irrespective of $D$. The spin-glass solutions can be found
numerically. We have used an iterative approach to search for such solutions by
rewriting the magnetization equations (\ref{tap0}) as
$m_{i,n+1}= {\cal F}(\{m_{i,n}\})$. The criterion adopted for convergence was

\begin{equation}
{1\over N}\sum_{i=1}^{N} |m_{i,n+1}-m_{i,n}| < 10^{-6} ~.
\end{equation}
In the present case this
method works finely and allows us to obtain as many spin-glass
solutions as we could. That will not be so if we were working in a non-zero
temperature regime as happens in the SK model \cite{BrayMoore79}.
We were thus able to improve the results obtained earlier \cite{Rehker99},
analysing systems varying from a few spins up to $N=1000$ spins. We
have also varied the number of realizations of random interactions, $N_R$.
Within each realization, the number of samples, $N_S$, was determined
from distinct initial conditions obtained as spin-glass solutions of the
TAP equations for $D=0$. For a given sample, we determine the solution with
lowest energy density $f_{\rm min}$ at $D=0$ and keep only solutions
with energy densities such that$|f/f_{\rm min}-1| < .05$. The solutions
thus kept, typically less than 5\% of $N_S$, are then used as initial
conditions to upgrade the solutions for a new value of $D$. This method
allows us to obtain the spin-glass energy density as a function of $D$
for each surviving sample. Finally, we averaged over the number of surviving
samples and over the number of realizations. Fig. 1 summarizes our
findings for the energy density. The error bars for the energy
density are comparable in size to the symbols used on that figure, which
give us much confidence on our results. We have also verified that for
small systems (up to $N=200$), the present method reproduces the results
presented in Ref. \cite{Rehker99}, but with strong fluctuations. Thus,
we find that the first-order transition which occurs when the spin-glass
energy density becomes zero is located at

\begin{equation}
                   D/J = 0.858\pm0.008 ~.
\end{equation}

This result improves the one found previously \cite{Rehker99}.
Nevertheless, it is also remarkably different from those obtained
within the replica approach \cite{daCosta94,Feldmann99,YokoiU}. We have
no sound explanation for this discrepancy between two seemingly
equivalent methods as TAP formulation and replica treatment. We hope
that other methods such as Monte Carlo simulation, or exact
determination of the ground state for finite systems, could help us in
determining the zero-temperature transition in a definite way.

The total number of solutions to Eq. (\ref{tap0}) can also be computed using
the methods introduced by De Dominicis et al. \cite{DeDominicis80} or
Bray and Moore \cite{BrayMoore80}, and can be shown to give the same results.
The latter method will be used in this note. Let us rewrite the magnetization
equations as

\begin{equation}
m_i = \varphi(h_i) ~ , \nonumber
\end{equation}
where $\varphi(h_i) = {\rm sgn}(h_i)\Theta(|h_i|-D)$. We also need to
introduce the Edwards-Anderson order parameter defined as

\begin{equation}
q = \frac{1}{N} \sum_{i=1}^{N} m_{i}^{2} ~ .
\end{equation}
Hence the total number of solutions $\left< N_{s} \right>$ is given by

\begin{eqnarray}
\left< N_{s}\right> &=& N\int_{0}^{1}dq \int_{-1}^{+1}\prod_{i}dm_{i}
\int_{-\infty}^{+\infty}\prod_{i}dh_{i} \delta(Nq-\sum_{i}m_{i}^{2}) \nonumber \\
& & \prod_{i}\left[ \delta(m_{i}-\varphi(h_{i})) \left< \delta(h_{i}-\sum_{j\ne i}
J_{ij}m_{j}) \right>\right] ~ ,
\end{eqnarray}
where $\left< {\cal O}\right>$ means the average of ${\cal O}$ over the
random bonds. Introducing integral representations for the delta functions
involving $q$ and $h_{i}$ in the above expression gives

\begin{eqnarray}
\left< N_{s} \right> &= &N\int_{0}^{1}dq \int_{-i\infty}^{+i\infty}{d\lambda \over 2\pi i}
\int_{-1}^{+1}\prod_{i}dm_{i}
\int_{-\infty}^{+\infty}\prod_{i}dh_{i}\int_{-\infty}^{+\infty}
\prod_{i}{dy_{i}\over \sqrt{2\pi}}  \nonumber \\
& \times &\prod_{i}\delta(m_{i}-\varphi(h_{i})) \exp{\left(N\lambda q -
\lambda \sum_{i}m_{i}^{2} + i\sum_{i}y_{i}h_{i} -
\frac{q}{2N}\sum_{i}y_{i}^{2}  \right)} \nonumber \\
& \times & \left< \exp\left(-i\sum_{(ij)}J_{ij}(y_{i}m_{j}+y_{j}m_{i})^{2}\right)
\right> ~. \label{ns1}
\end{eqnarray}
Performing the average over the Gaussian bond distribution, the last
factor in (\ref{ns1}) becomes

\begin{equation}
\exp\left[-\frac{J^2}{2N}\sum_{(ij)}(y_{i}m_{j}+y_{j}m_{i})^{2}\right]
= \exp\left[- \frac{J^2}{2N}q\sum_{i}y_{i}^{2}
- \frac{J^2}{2N}\left(\sum_{i}y_{i}m_{i}\right)^2\right] ~,
\label{ep1}
\end{equation}
where we have neglected terms that do not contribute in the
thermodynamic limit. The final factor in (\ref{ep1}) is simplified
using the identity

\begin{equation}
{\rm e}^{\displaystyle{-\frac{J^2}{2N}(\sum_{i}y_{i}m_{i})^2}} =
\int_{-\infty}^{+\infty}{dt\over \sqrt{2\pi/N}}
{\rm e}^{\displaystyle{-{N\over 2}t^2+iJt\sum_{i}y_{i}m_{i}}} ~. \label{ep2}
\end{equation}
Assembling the results (\ref{ep1}) and (\ref{ep2}) into (\ref{ns1}), we
obtain

\begin{eqnarray}
\left<  N_{s} \right> &=& N^{3/2}\int_{0}^{1}dq \int_{-\infty}^{+\infty}{dt\over \sqrt{2\pi}}
\int_{-i\infty}^{+i\infty}{d\lambda \over 2\pi i} \int_{-1}^{+1}\prod_{i}dm_{i}
\int_{-\infty}^{+\infty}\prod_{i}dh_{i}\int_{-\infty}^{+\infty}
\prod_{i}{dy_{i}\over \sqrt{2\pi}}
\nonumber \\
& \times &
\prod_{i}\delta(m_{i}-\varphi(h_{i})) \nonumber \\
& \times &
\exp{\left[-{N\over 2}t^2 + N\lambda q - \lambda \sum_{i}m_{i}^{2}
- \frac{J^2q}{2N}\sum_{i}y_{i}^{2} + i\sum_{i}(h_{i}+Jtm_{i})y_{i}\right]} ~.
\end{eqnarray}
In the thermodynamic limit, the above expression is dominated by the
saddle point of the integrand with respect to the variables $t$,
$\lambda$ and $q$. Thus, we have

\begin{equation}
\left< N_{s} \right> \approx \exp{(N\phi_{T})}
\end{equation}
where $\phi_{T}$ is the saddle point of

\begin{equation}
\phi = -\frac{1}{2}t^{2} + \lambda q + \ln{\Xi} ~,
\end{equation}
and $\Xi$ is given by
\begin{equation}
\Xi =  2 \int_{-D/J}^{0}{dx \over \sqrt{2\pi}} {\rm e}^{-x^2/2q}
       + 2 \int_{-\infty}^{D/J}{dx \over \sqrt{2\pi q}}
       {\rm e}^{-(x-Jt)^2/2q- \lambda} ~,
\end{equation}
for $D>0$. For $D\le 0$ we recover the result for the
SK model which gives $\phi_{T} \approx 0.1992$
\cite{DeDominicis80,BrayMoore80,Tanaka80}. Therefore, we only need to
determine $\phi_{T}$ for positive values of $D$. This is achieved
numerically solving the saddle point equations for $t$,
$\lambda$ and $q$. The result is presented in figure 2. It is
interesting to note that $\phi_{T}$ has a smooth behavior as a
function of $D$, increasing from $0.1992$, reaching a maximum around
$D\approx0.550J$ and then decreasing continuously to zero at
$D\approx1.225J$. The maximum is attained when the spin-glass phase
presents a large number of spins in the $S=0$ state, whereas the
remaining spins occupying the $S=\pm1$ states are still in conflict due to
frustration and randomness. Thus, the spin-glass phase may become
more complex for intermediate values of $D$. As this parameter
increases still further, eventually more and more spins prefer to stay
in the $S=0$ state and finally the paramagnetic phase becomes the unique
stable phase. It is also interesting to note that in the region where the
first-order transtion is expected to occur the number of solutions is
almost the same as in the Sherrington-Kirkpatrick model.

\vskip \baselineskip
\noindent
{\large\bf Acknowledgments}

\noindent
One of us (FAC) would like to thank E.M.F. Curado for his kind
hospitality at the Centro Brasileiro de Pesquisas F\'{\i}sicas.
We also thank F.D. Nobre and P.M.C. de Oliveira for their comments
and suggestions.

\newpage

\centerline{{\large\bf Figure Captions}}

\vskip 2\baselineskip
\noindent
{\bf Fig. 1.} Spin-glass energy density $f$ as a function of $D/J$
for $N = 300 ~(\triangle), 400 ~(\Box), 500 ~(\diamond)$ and $600 ~(\circ)$
spins. The first-order transition is located when the curves cross
the horizontal zero axis, since the paramagnetic energy density is zero
for any value of $D$.

\vskip 2\baselineskip
\noindent
{\bf Fig. 2.}  The logarithm of the total number of TAP solutions per spin,
$\phi_{T} = N^{-1}\ln\left<N_s\right>$, as a function of $D/J$.

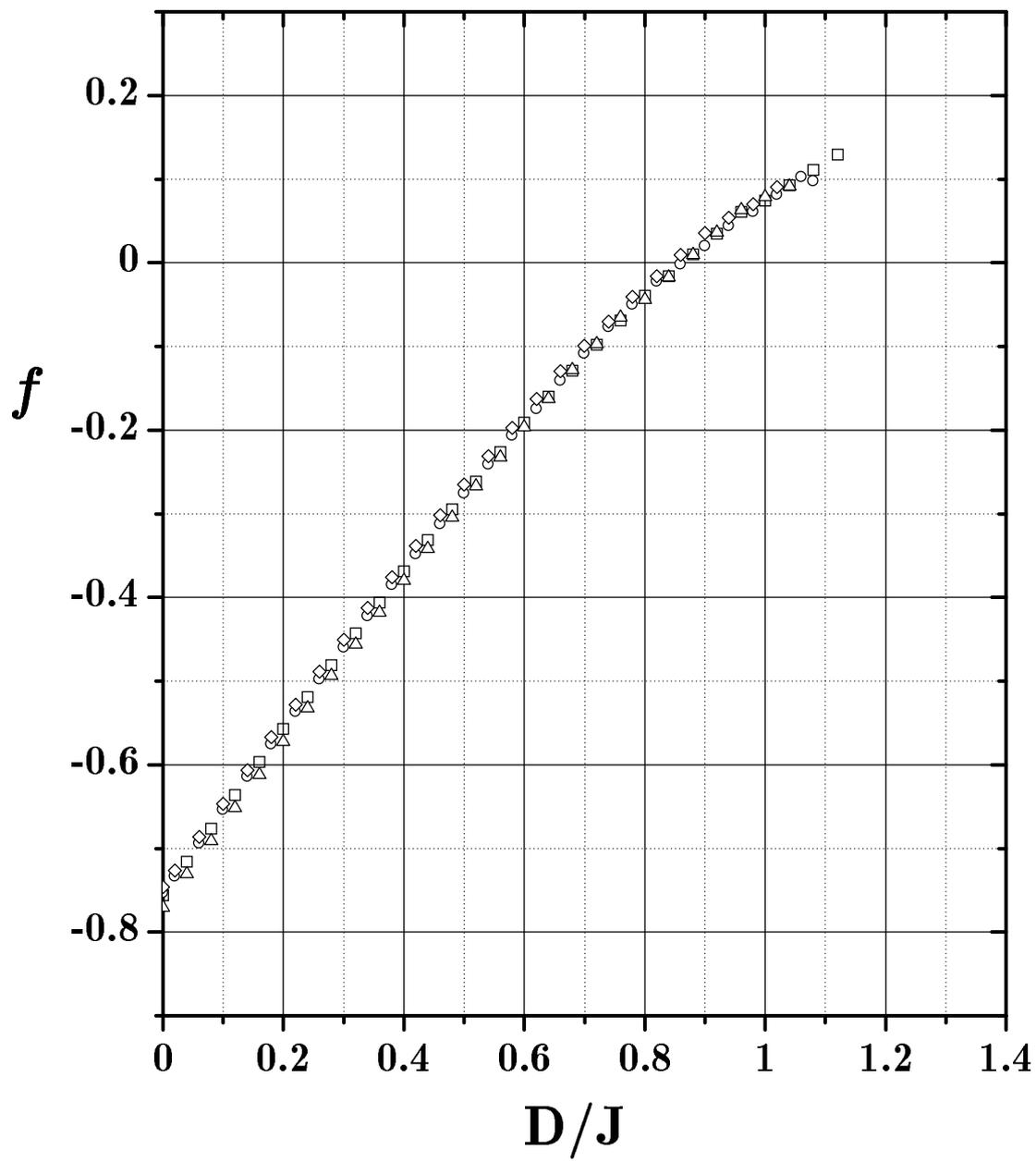
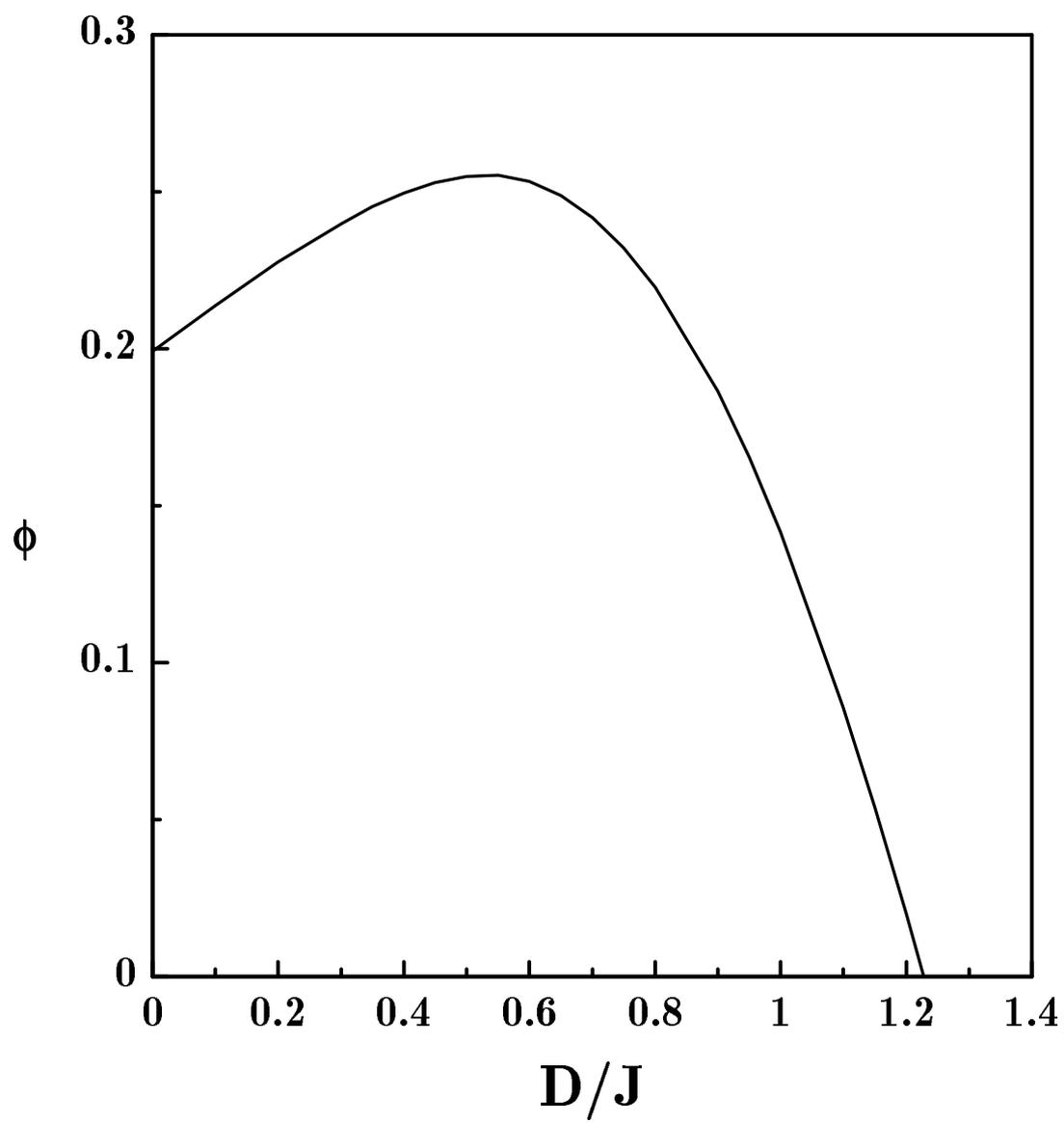
\end{document}